# Electronic and Optical Properties of γ- and θ- Alumina by First Principle Calculations

Ahmed S. Jbara[1, 2, 3, *], Zulkafli Othaman[1, 3], H. A. Rahnamaye Aliabad[4],

and M. A. Saeed[3, 5]

[1] *Center for Sustainable Nanomaterials, Universiti Teknologi Malaysia, Skudai-81310, Johor Bahru, Malaysia*

[2] *Physics Department, Science College, Al-Muthanna University, Samawah- 66001, Iraq*

[3] *Department of Physics, Faculty of Science, Universiti Teknologi Malaysia, Skudai-81310, Johor Bahru, Malaysia*

[4] *Department of Physics, Hakim Sabzevari University, Sabzevar-8855192301, Iran*

[5] *Division of Science and Technology, University of Education, Township, Lahore- 54770, Pakistan*

*\*Corresponding author email: ahmedsbhe@yahoo.com*

**Abstract**: The electronic and optical parameters of $\gamma$-$Al_2O_3$ and $\theta$-$Al_2O_3$ have been studied by using the first principle within the framework of density function theory (DFT). The computational approach is based on full-potential linearized augmented plane wave method (FP-LAPW) within the generalized gradient approximation (GGA), local density approximation (LDA), and modified Becke-Johnson potential (mBJ). The results show that these compounds have a direct gap ($\Gamma$-$\Gamma$) of about 5.375 eV and 4.716 eV for $\gamma$-$Al_2O_3$ and $\theta$-$Al_2O_3$, respectively. Several optical parameters of these materials are also investigated. The values of the real part of dielectric constant are found to be 3.259 and 3.694 for $\gamma$-$Al_2O_3$ and $\theta$-$Al_2O_3$, respectively, which are close to the experimental one (3.416). The refractive index is 1.806 and 1.922 for $\gamma$-$Al_2O_3$ and $\theta$-$Al_2O_3$ respectively, and shows a good agreement with the experimental result which is 1.86. GGA findings are consistent with the experimental results and are better than the other approximations. There are no salient differences between GGA and LDA results. The results advocate using this material as a transparent conducting layer in solar cell structure, which can be operated in a wide energy range.

1. Introduction

Currently, great challenges and opportunities in the industry of solar cells are mainly focused on incorporating nanostructure materials in their structure, which lie in making substantial improvements in these materials to increase efficiency for the generation, conversion, transmission and use of energy. In recent years, aluminium oxide (alumina) $Al_2O_3$ material has been instrumental in overcoming these challenges [1-4].

Alumina has attracted much attention among single metal oxide because of the high specific surface area and a large number of defects in its crystalline structure. They are widely used in ceramic applications such as microporous catalysts, ultra-hard coatings, abrasives and polishing [5, 6], and electroluminescent flat-screen displays [7]. The nanoparticles' alumina are also used as fillers for ceramic matrix composite materials [8] and adsorbent to remove crystal violet [9] and phenol [10, 11]. Based on literature, there are seven phases of alumina, also known as "Transition Aluminas", such as chi-, kappa-, gamma-, theta-, delta-, eta-, and rho- alumina. Although corundum structure, also called alpha-alumina, is not a transition alumina, it is the last crystalline structure formed by increasing the temperature on the transition alumina. All these transition aluminas are metastable having oxygen sub-lattice that is near to close-packed (either face-centered cubic or hexagonal close-packed) [12]. Gamma- $Al_2O_3$ ($\gamma$-$Al_2O_3$) with space group *Fd3m* [12, 13], was the first classified as a transition alumina phase by Ulrich [14]. The gamma prefix has been used for the structure obtained from the dehydration sequence of boehmite at 400 ºC. Various studies have shown that theta phase (θ-$Al_2O_3$) appears at high temperature [15] around 1000 ºC, which has a monoclinic crystal structure with *C12/m1* space group.

At present, approaches based on density functional theory (DFT) are mostly used to calculate the physical properties such as structural, electronic and optical properties. The computational approaches are considered more and more popular in many fields such as material science, condensed matter, and quantum chemistry. In some cases, it has even replaced the experimental method, such as when they are difficult to execute under standard conditions, or to understand the behavior of physics phenomena for real materials and to make specific predictions of new materials. The computational method also reduces the time and cost. Various computer packages employed DFT in quantum chemistry and solid state physics software packages, mostly along with other methods, such as WIEN2k code [16].

The main part of DFT approach is the structured data, where there is no perfect crystalline structure for both alumina phases. Therefore, most researchers take the nearest structure of materials and then modify it [17-20]. For instance, the structure and bonding in θ-$Al_2O_3$ have been studied by means of DFT method with local density approximation (LDA), and the outstanding result of this work is the band gap value around 4.64 eV [20]. Cai, *et al.* [21] have investigated theoretical modelling for the dehydration mechanism of the transformation between γ-$Al_2O_3$ and θ-$Al_2O_3$ by means of DFT as implemented in the Cambridge Serial Total Energy Package (CASTEP). Vienna *ab initio* simulation program VASP in the framework of DFT has been used to compute electronic structure of γ-$Al_2O_3$ with generalized gradient approximation (GGA) and LDA as the exchange-correlation potential performed by Menéndez-Proupin and Gutiérrez [22]. They have suggested spinel and non-spinel model for γ-$Al_2O_3$. The results have shown that there is no change in the band structure of both models and the band gap value is found be 3.96 eV. Paglia, *et al.* [13] have shown an extensive computational study of γ-Al$_2$O$_3$ by means of SIESTA code, which include the geometric analysis of around 1.47 billion spinel structures, followed by derivative method energy minimization calculations of nearly 122000 structures. This research has contributed further insight on spinel and non-spinel structural models. Authors believed that simulation was more favorable with a structures non-spinel site occupancy, while later, the spinel model was proposed [23]. In this work, structure files of γ- and θ-Al$_2$O$_3$ are created and then used to calculate the electronic and optical properties of these materials.

## 2. Computational Description

The electronic and optical properties of these crystal structures are investigated using the highly accurate full-potential linearized augmented plane wave (FP-LAPW) within the framework of DFT, as implemented in the WIEN2k computer package [24, 25]. Generalized gradient approximation (GGA), local density approximation (LDA), and modified Becke–Johnson potential (mBJ) are used for the exchange-correlation interaction [26, 27]. GGA is an extension or a next logical step to improve LDA potential, which includes some knowledge of in-homogeneities in the density by making the exchange-correlation contribution of every infinitesimal volume not only depending on the local density in that volume, but also in the density in the neighbouring volumes. According to the previous studies, the GAA predictions are the same if not better than the LDA for getting structural [28], electronic [29], and optical [30] properties of materials. Therefore, the electronic and optical properties of both alumina phases have been computed under the basic approximations of GGA and LDA to show the little differences between their results.

The calculation starts by creating structure files for both alumina phases depending on available information from previous experimental works (see Figure 1). In this calculation, the separation energy between the core and valence states is -8.0 Ry. In the full potential scheme, the wave functions inside the atomic spheres are expanded in terms of spherical harmonics up to $l_{max} = 10$, and in terms of plane waves with a cut-off $R_{MT} K_{MAX} = 7$, where $R_{MT}$ denotes the smallest muffin-tin radius, and $K_{MAX}$ gives the magnitude of the largest K vector in the plane wave expansion. 50 k-points are used in the Brillouin zone integration. The self-consistency is assumed to be attained when the total energy difference between successive iterations is less than $10^{-5}$ Ry.

3. Results and Discussion

The calculated band structures for both phases of alumina are shown in Figure 2. The Fermi level is set at zero on the top of the valence band. It is clear from the figures obtained using GGA and mBJ that both structures show direct ($\Gamma$-$\Gamma$) band gap. Table 1 shows the computed values compared with the experimental results, along with band gap values obtained from other theoretical works. The band gaps obtained using GGA are close to the experimental results, but the values obtained using mBJ are larger than the experimental values. The same behavior is also observed for other compounds [27]. These band gap values are also better than the previously obtained theoretical values. It is very apparent from Table 1 that there are no salient differences between GGA and LDA, with differences of about 0.305 eV and 0.117 eV for $\gamma$-$Al_2O_3$ and $\theta$-$Al_2O_3$, respectively. Also, Figure 2 illustrates total density of states (DOS) for $\gamma$-$Al_2O_3$ and $\theta$-$Al_2O_3$ by using GGA and mBJ potential. The dashed line at 0 eV represents Fermi level $E_F$. From the comparison, one can see that all states in the conduction band are shifted toward high energy as we move from GGA to mBJ approaches, which causes increasing in the band gap.

In order to explain the differences between GGA and mBJ approximations in the optical properties, the different optical spectra such as real $\varepsilon_1(\omega)$ and imaginary parts $\varepsilon_2(\omega)$ of the dielectric function, reflectivity $R(\omega)$, refractive index $n(\omega)$, extinction coefficient $k(\omega)$, and absorption coefficient $\alpha(\omega)$ that are frequency dependent are plotted.

The real part of the dielectric function $\varepsilon_1(\omega)$ gives information about the electronic polarizability of material. The static dielectric function is inversely dependent on the band gap values. The comparison of the static dielectric constant $\varepsilon_1(0)$ under GGA and mBJ for both alumina phases are given in Table 2, and for comparison with the experimental results, $\varepsilon_1(\omega)$ is also estimated at 5 eV corresponding energy. One can clearly see in this table a convergence between GGA and LDA values for both γ-Al$_2$O$_3$ and θ-Al$_2$O$_3$. The interesting feature of the $\varepsilon_2(\omega)$ curve is its first critical point, also known as optical absorption edge. Figure 3 reveals that the optical absorption edge is found at 5.293 eV (7.197 eV) and 4.612 eV (7.046 eV) with GGA (mBJ), for γ-Al$_2$O$_3$ and θ-Al$_2$O$_3$, respectively. The shift in the calculated spectra of absorption edge is in a good agreement with those from band structure and corresponding DOS of both alumina phases. The various peaks of this curve are also signified to different interband transitions between the conduction and valence band states. The high energy shift with mBJ approximation has been found for the real and imaginary part of the dielectric function, and the same trend can be clearly observed within γ-Al$_2$O$_3$ and θ-Al$_2$O$_3$. Due to unavailability of experimental value of the dielectric function for γ- and θ-Al$_2$O$_3$, only a comparison with α-Al$_2$O$_3$ results is provided to give an indication of the accuracy of our theoretical predictions for the γ-and θ- phases. Figure 4 shows the previously reported experimental data of α-Al$_2$O$_3$ [31], and it is very similar in the general behaviour between these spectra and the spectra obtained in this work (Figure 3).

After getting the real and imaginary parts of the dielectric function, it becomes easy to calculate the reflectivity $R(\omega)$, refractive index $n(\omega)$, extinction coefficient $k(\omega)$, and absorption coefficient $\alpha(\omega)$. The frequency dependence of reflectivity $R(\omega)$ for both alumina phases is shown in Figure 5. There is no considerable anisotropy between spectral under GGA and mBJ except the energy shift to higher values with mBJ. The same behaviour has been observed in the previous optical parameters. It is important to note that reflectance occurs along the spectrum regions 13 eV - 25 eV (15 eV - 27 eV) and 16 eV - 25.5 eV (19 eV - 26 eV) in case of GGA (mBJ) for γ-and θ-Al$_2$O$_3$, respectively. These regions exactly correspond to the negative values of $\varepsilon_1(\omega)$, which refer to reflected incident waves from a material. The magnitude of the reflectivity at zero frequency is listed in Table 2, while the maximum value of reflectivity appears at 20.068 eV (22.109 eV) and 23.171 eV (23.606 eV) GGA (mBJ) for γ-Al$_2$O$_3$ and θ-Al$_2$O$_3$, respectively. Figure 5 also shows α-Al$_2$O$_3$ reflectance spectrum that is plotted based on the experimental measurements from earlier work [32]. One can see matching in the energy regions mentioned above.

In industry, it is necessary to know the refractive index of materials to accurately design the optical devices. The refractive index $n(\omega)$ and extinction coefficient $k(\omega)$ curves for γ- and θ-Al$_2$O$_3$ are illustrated in Figure 6, while for α-Al$_2$O$_3$, experimental results of Arakawa and Williams [31] and French, *et al.* [32] are put in Figure 7 to make a comparison. At first glance, one can observe an excellent agreement with the experiment. Minor differences in spectra are mainly attributed to different phases. The calculated static values of refractive index $n(0)$ with both approximations GGA and mBJ are given in Table 2. The present theoretical values of the refractive index at 5 eV are 1.806 (1.631) and 1.922 (1.555) computed with GGA (mBJ) for γ- and θ-Al$_2$O$_3$, respectively. These values are close to the experimental value of 1.86 [31]. As we know, that the refractive index spectrum is similar to the real part of the dielectric function ($n(\omega) = \sqrt{\varepsilon_1(\omega)}$) [33], therefore we attained this fact here (see Figures 3 and 6). Some applications require the highest value of refractive index and with corresponding energies, and the present results for γ-Al$_2$O$_3$ with a case of refractive index (corresponding energy) are 2.528 (9.374 eV) and 2.224 (11.143 eV) for GGA and mBJ, respectively. In the case of θ-Al$_2$O$_3$, the values are 2.172 (8.449 eV) and 1.888 (10.925 eV) with both GGA and mBJ potential, respectively. The trend of similarity between GGA and LDA results are also observed in the $R(\omega)$ and $n(\omega)$ values (see Table 2).

The extinction coefficient $k(\omega)$ is also calculated within GGA and mBJ for both alumina phases under consideration and is shown in Figure 6. This is evidence that the overall behaviour of $k(\omega)$ is close to that of the imaginary part of the dielectric functions $\varepsilon_2(\omega)$ (see Figure 3), which affirms the established theory. Of course, their tendency in detail is not precisely identical since the extinction coefficients depend on both the real and imaginary parts of the dielectric functions according to [33].

Absorption coefficient $\alpha(\omega)$ as a function of frequency for γ- and θ-$Al_2O_3$ is shown in Figure 8 (a) and (b), respectively. The blue shift in the spectra is also shown with mBJ approximation for both phases. It is clearly seen from these figures that the spectra of absorption is stable with low values up to the value of band gap then suddenly rises, which confirms the band gap values that have been obtained previously from the band structure. The variation peaks in the strong absorption regions appear around 8 eV - 32 eV (10 eV - 34 eV) and 5 eV - 26 eV (7.5 eV - 26 eV) in a case of GGA (mBJ) for γ- and θ-$Al_2O_3$, respectively, whereas the present results refer to the maximum values of $\alpha(\omega)$ ranged from 274×10$^4$ cm$^{-1}$ (288×10$^4$ cm$^{-1}$) and 244×10$^4$ cm$^{-1}$ (266×10$^4$ cm$^{-1}$) for γ- and θ-$Al_2O_3$, respectively. Absorption regions are stretching more to the high energy, while almost non-existent in the visible range ($\alpha(\omega) \approx 0$), meaning that the materials are transparent to visible wavelengths. All these data show an increase in the prospect of using these materials in optoelectronic devices that can be operated on a wide range of ultra-violet (UV) spectrum.

## 4. Conclusion

In this work, FP-LAPW method has been used to investigate the electronic and optical properties of γ-$Al_2O_3$ and θ-$Al_2O_3$ by applying different approximations such as GGA, LDA, and mBJ. Band structures show direct band gap (Γ-Γ) for both γ-Al$_2$O$_3$ and θ-Al$_2$O$_3$ having values about 5.375 eV and 4.646 eV, respectively. GGA findings are consistent with the previous experimental results and these results are better than the previous theoretical values. Also, these results reveal that there are no salient differences between GGA and LDA results. The refractive index $n(5)$ values have achieved a good agreement with the experimental result (1.86), where our results are referred to 1.806 (1.922) with a case of γ-Al$_2$O$_3$ (θ-Al$_2$O$_3$). Optical absorption spectra indicate a possibility of using these materials as a filter for various energies in the UV spectrum.

**Acknowledgment**

One of the authors (Ahmed S. Jbara) is grateful to the Ministry of Higher Education and Scientific Research (MOHESR) of Iraq for providing research grant. Authors would like to acknowledge Universiti Teknologi Malaysia (UTM) and Ministry of Higher Education of Malaysia for providing the necessary facilities (Vote no. 12-H75/4F736) and Center for Sustainable Nanomaterials and Department of Physics, UTM for technical support.

**References**

1. C.-M. Lung, W.-C. Wang, C.-H. Chen, L.-Y. Chen, and M.-J. Chen, "ZnO/Al2O3 core/shell nanorods array as excellent anti-reflection layers on silicon solar cells", Materials Chemistry and Physics **180**, p. 195-202 (2016).

2. W.-P. Goh, E. L. Williams, R.-B. Yang, W.-S. Koh, S. Mhaisalkar, and Z.-E. Ooi, "Optimal Shell Thickness of Metal@Insulator Nanoparticles for Net Enhancement of Photogenerated Polarons in P3HT Films", ACS Applied Materials & Interfaces **8**(4), p. 2464-2469 (2016).

3. Z. K. Heiba, N. G. Imam, and M. Bakr Mohamed, "Structural optical correlated properties of SnO2/Al2O3 core@ shell heterostructure", Journal of Molecular Structure **1115**, p. 156-160 (2016).

4. J. Keller, F. Gustavsson, L. Stolt, M. Edoff, and T. Törndahl, "On the beneficial effect of Al2O3 front contact passivation in Cu(In,Ga)Se2 solar cells", Sol. Energ. Mat. Sol. C. **159**, p. 189-196 (2017).

5. T. Hatanaka, T. Hiraiwa, M. Matsukura, N. Aoki, and S. Imai, *Alumina-based ceramics materials, abrasive materials and method for the manufacture of the same*. 1993, Google Patents.

6. Y. Bhatnagar, R. T. Rozbicki, and R. Mulpuri, *Optical device fabrication*. 2012, Google Patents.

7. R. Doremus, *Alumina*, in *Ceramic and Glass Materials*, J. Shackelford and R. Doremus, Editors. 2008, Springer US. p. 1-26.

8. M. Vahtrus, M. Umalas, B. Polyakov, L. Dorogin, R. Saar, M. Tamme, K. Saal, R. Lõhmus, and S. Vlassov, "Mechanical and structural characterizations of gamma- and alpha-alumina nanofibers", Materials Characterization **107**, p. 119-124 (2015).

9.  A. Adak, M. Bandyopadhyay, and A. Pal, "Adsorption of anionic surfactant on alumina and reuse of the surfactant-modified alumina for the removal of crystal violet from aquatic environment", Journal of Environmental Science and Health **40**(1), p. 167-182 (2005).

10. A. Adak and A. Pal, "Removal of phenol from aquatic environment by SDS-modified alumina: Batch and fixed bed studies", Separation and Purification Technology **50**(2), p. 256-262 (2006).

11. M. C. Patterson, N. D. Keilbart, L. W. Kiruri, C. A. Thibodeaux, S. Lomnicki, R. L. Kurtz, E. D. Poliakoff, B. Dellinger, and P. T. Sprunger, "EPFR formation from phenol adsorption on Al2O3 and TiO2: EPR and EELS studies", Chem. Phys. **422**, p. 277-282 (2013).

12. I. Levin and D. Brandon, "Metastable Alumina Polymorphs: Crystal Structures and Transition Sequences", Journal of the American Ceramic Society **81**(8), p. 1995-2012 (1998).

13. G. Paglia, A. L. Rohl, C. E. Buckley, and J. D. Gale, "Determination of the structure of γ-alumina from interatomic potential and first-principles calculations: The requirement of significant numbers of nonspinel positions to achieve an accurate structural model", Phys. Rev. B **71**(22), p. 224115 (2005).

14. F. Ulrich, "Notiz über die Kristallstruktur der Korund-Hämatitgruppe", Norsk Geol. Tidsskr **8**, p. 113 (1926).

15. G. Busca, "The surface of transitional aluminas: A critical review", Catalysis Today **226**, p. 2-13 (2014).

16. P. Blaha, K. Schwarz, G. Madsen, D. Kvasnicka, and J. Luitz, "WIEN2k, An Augmented Plane Wave+ Local Orbitals Program for Calculating Crystal Properties", (2001), ISBN 3-9501031-1-2: Vienna University of Technology, Austria.

17. H. P. Pinto, R. M. Nieminen, and S. D. Elliott, "Ab initio study of gamma-Al2O3 surfaces", Phys. Rev. B **70**(12), p. 125402 (2004).

18. M. Yazdanmehr, S. Asadabadi, A. Nourmohammadi, M. Ghasemzadeh, and M. Rezvanian, "Electronic structure and bandgap of γ-Al2O3 compound using mBJ exchange potential", Nanoscale Res Lett **7**(1), p. 1-10 (2012).

19. L. Fu and H. Yang, "Tailoring the Electronic Structure of Mesoporous Spinel γ-Al2O3 at Atomic Level: Cu-Doped Case", The Journal of Physical Chemistry C **118**(26), p. 14299-14315 (2014).

20. S.-D. Mo and W. Ching, "Electronic and optical properties of $\theta-Al_2O_3$ and comparison to $\alpha-Al_2O_3$", Phys. Rev. B **57**(24), p. 15219 (1998).

21. S.-H. Cai, S. N. Rashkeev, S. T. Pantelides, and K. Sohlberg, "Phase transformation mechanism between $\gamma$- and $\theta$-alumina", Phys. Rev. B **67**(22), p. 224104 (2003).

22. E. Menéndez-Proupin and G. Gutiérrez, "Electronic properties of bulk $\gamma-Al_2O_3$", Phys. Rev. B **72**(3), p. 035116 (2005).

23. A. R. Ferreira, E. Küçükbenli, A. A. Leitão, and S. de Gironcoli, "Ab initio 27 Al NMR chemical shifts and quadrupolar parameters for Al 2 O 3 phases and their precursors", Phys. Rev. B **84**(23), p. 235119 (2011).

24. K. Schwarz and P. Blaha, "Solid state calculations using WIEN2k", Computational Materials Science **28**(2), p. 259-273 (2003).

25. K. Schwarz, "DFT calculations of solids with LAPW and WIEN2k", Journal of Solid State Chemistry **176**(2), p. 319-328 (2003).

26. J. P. Perdew, K. Burke, and M. Ernzerhof, "Generalized Gradient Approximation Made Simple", Phys. Rev. Lett. **77**(18), p. 3865-3868 (1996).

27. F. Tran and P. Blaha, "Accurate band gaps of semiconductors and insulators with a semilocal exchange-correlation potential", Phys. Rev. Lett. **102**(22), p. 226401 (2009).

28. Z. Wu, R. Cohen, and D. Singh, "Comparing the weighted density approximation with the LDA and GGA for ground-state properties of ferroelectric perovskites", Phys. Rev. B **70**(10), p. 104112 (2004).

29. C. Stampfl, W. Mannstadt, R. Asahi, and A. Freeman, "Electronic structure and physical properties of early transition metal mononitrides: Density-functional theory LDA, GGA, and screened-exchange LDA FLAPW calculations", Phys. Rev. B **63**(15), p. 155106 (2001).

30. M. Lazzeri, C. Attaccalite, L. Wirtz, and F. Mauri, "Impact of the electron-electron correlation on phonon dispersion: Failure of LDA and GGA DFT functionals in graphene and graphite", Phys. Rev. B **78**(8), p. 081406 (2008).

31. E. T. Arakawa and M. W. Williams, "Optical properties of aluminum oxide in the vacuum ultraviolet", Journal of Physics and Chemistry of Solids **29**(5), p. 735-744 (1968).

32. R. H. French, H. Müllejans, and D. J. Jones, "Optical Properties of Aluminum Oxide: Determined from Vacuum Ultraviolet and Electron Energy-Loss Spectroscopies", Journal of the American Ceramic Society **81**(10), p. 2549-2557 (1998).

33. A. M. Fox, *Optical properties of solids*. Vol. 3. 2001: Oxford University Press, USA.

34. J. Gangwar, B. K. Gupta, P. Kumar, S. K. Tripathi, and A. K. Srivastava, "Time-resolved and photoluminescence spectroscopy of [small theta]-Al2O3 nanowires for promising fast optical sensor applications", Dalton Transactions **43**(45), p. 17034-17043 (2014).

**Tables**

**Table 1:** The calculated band gaps (eV) by GGA, LDA and mBJ potential for γ-and θ-Al$_2$O$_3$ compared with the previous theoretical and experimental data.

| Compounds | Theoretical Results | | | | | Experimental results |
|---|---|---|---|---|---|---|
| | Present work | | | Other works | | |
| | GGA | LDA | mBJ | GGA | LDA | |
| γ-Al$_2$O$_3$ | 5.375 | 5.680 | 7.233 | 3.97[a] 4.13[b] 4.1[c] | | 6[c] |
| θ-Al$_2$O$_3$ | 4.716 | 4.599 | 7.063 | | 4.64[d] | 5.16[e] |

a: [17], b: [18], c: [19], d: [20], and e: [34]

**Table 2:** The significant optical parameters calculated by GGA, LDA and mBJ potential for γ- and θ-Al$_2$O$_3$ compared with the previous experimental results.

| Parameter | γ-Al$_2$O$_3$ | | | θ-Al$_2$O$_3$ | | | α-Al$_2$O$_3$ |
|---|---|---|---|---|---|---|---|
| | GGA | LDA | mBJ | GGA | LDA | mBJ | Experimental |
| $\varepsilon_1(0)$ | 3.748 | 3.596 | 4.024 | 2.958 | 3.050 | 2.228 | Not available |
| $\varepsilon_1(5)$ | 3.259 | 3.249 | 2.653 | 3.694 | 3.872 | 2.418 | 3.416[a] |
| $R(0)$ | 0.102 | 0.096 | 0.112 | 0.070 | 0.074 | 0.039 | Not available |
| $R(5)$ | 0.083 | 0.082 | 0.058 | 0.099 | 0.106 | 0.047 | 0.086[b] |
| $n(0)$ | 1.936 | 1.896 | 2.006 | 1.720 | 1.747 | 1.493 | 1.76[b] |
| $n(5)$ | 1.806 | 1.803 | 1.631 | 1.922 | 1.968 | 1.555 | 1.86[a] |

a: [31], b: [32]

**Figures**

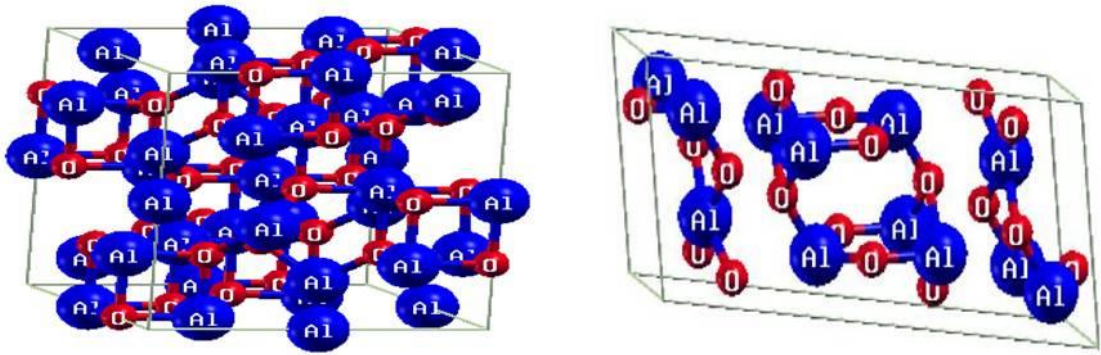

**Figure 1:** Crystal structure of γ-Al$_2$O$_3$ (left) and θ-Al$_2$O$_3$ compounds (right). The blues represent Al atoms, and the reds represent O atoms

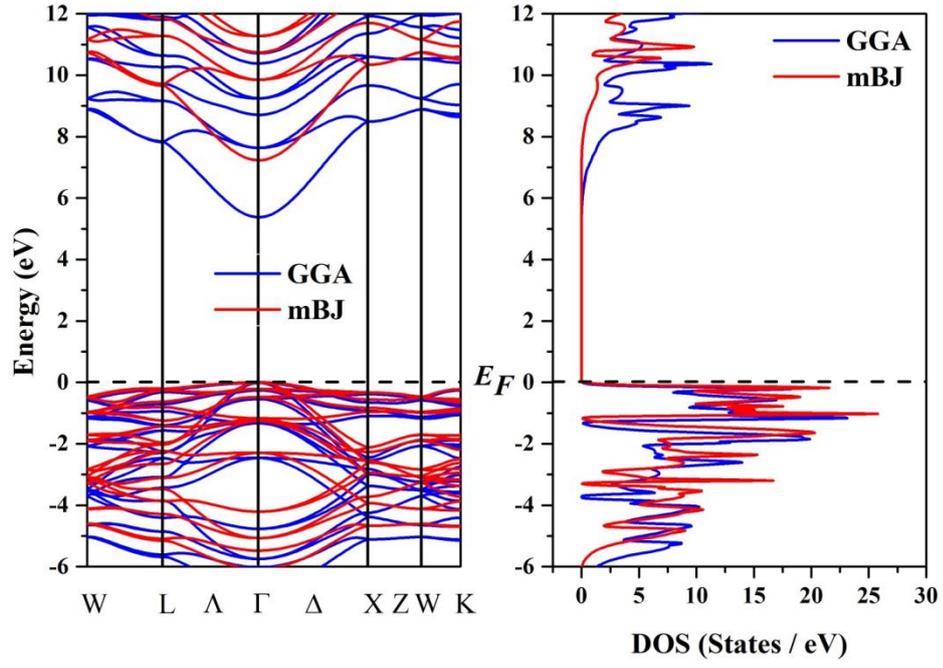

(a)

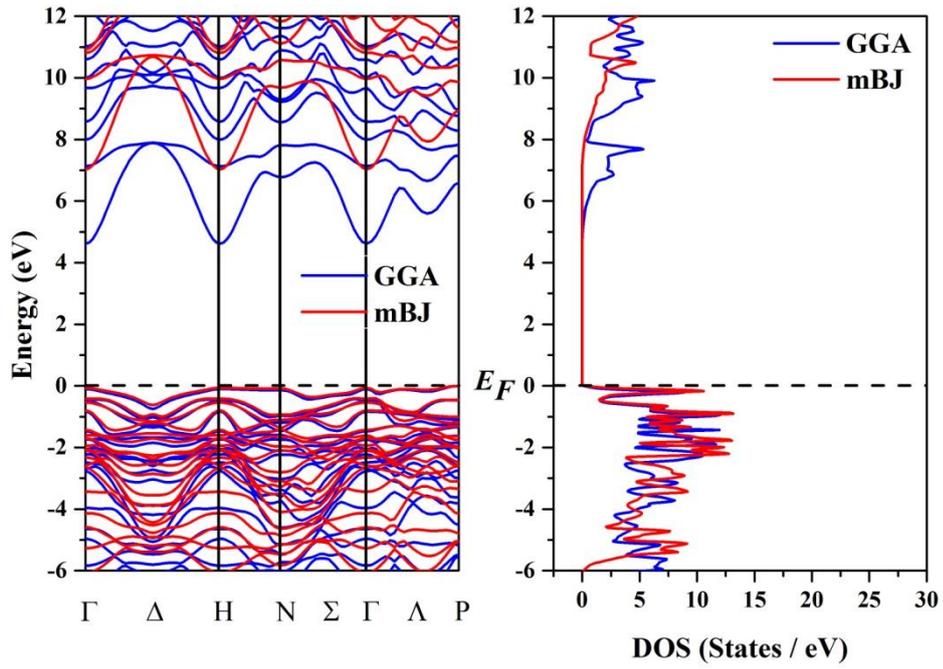

(b)

**Figure 2:** Band structure (left) and total DOS (right) for (a) γ-Al$_2$O$_3$ and (b) θ-Al$_2$O$_3$, obtained from GGA and mBJ potential.

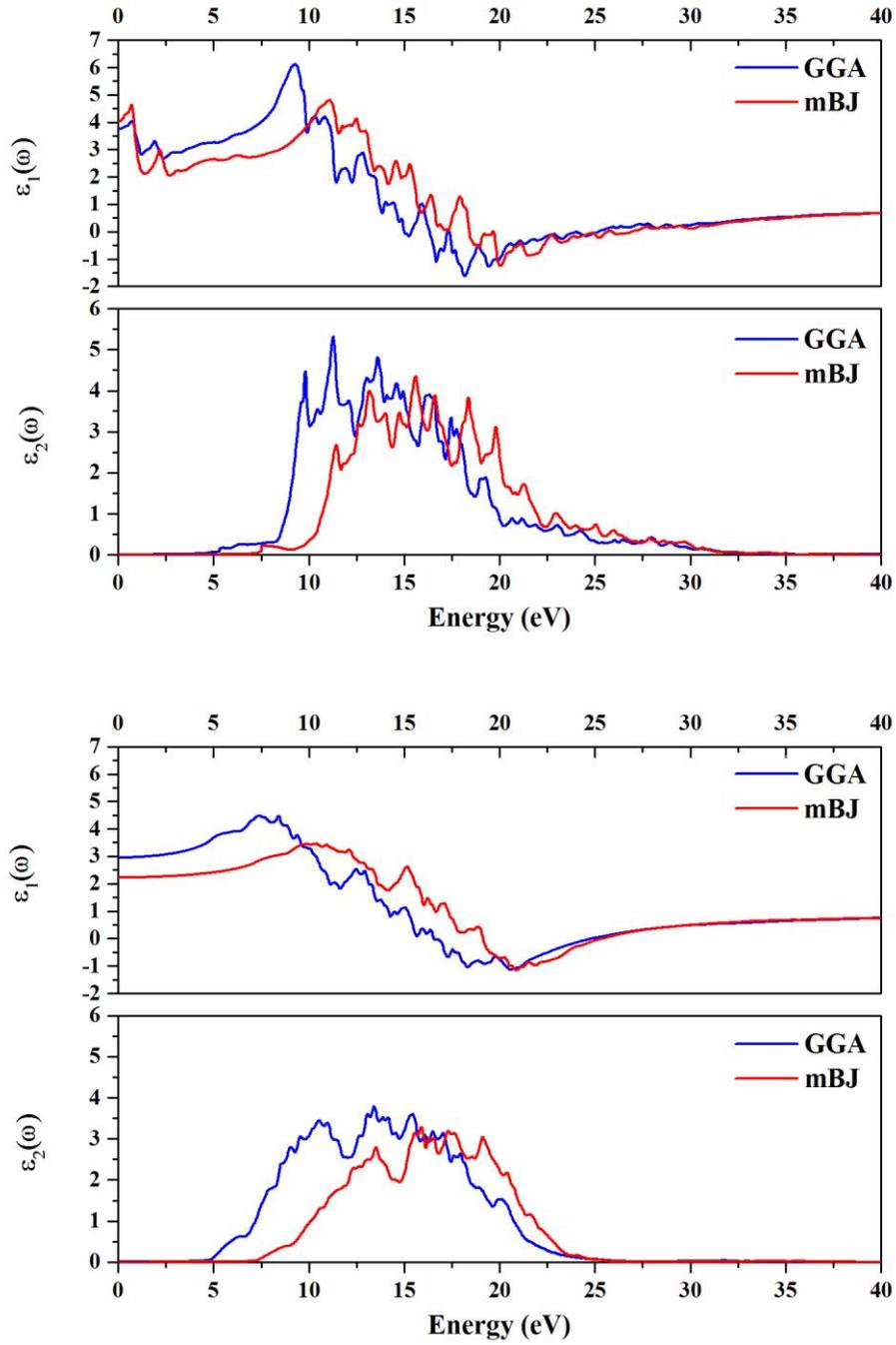

**Figure 3:** Real $\varepsilon_1(\omega)$ and imaginary $\varepsilon_2(\omega)$ part of the dielectric function for (a) $\gamma$-$Al_2O_3$ and (b) $\theta$-$Al_2O_3$ determined from GGA and mBJ potential.

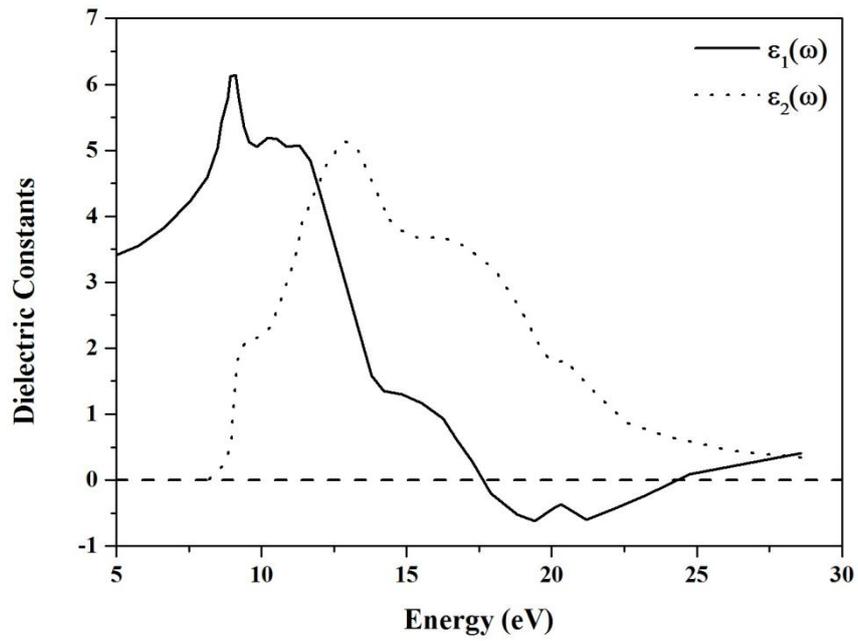

**Figure 4:** Experimental data for dielectric constants with real $\varepsilon_1(\omega)$ and imaginary $\varepsilon_2(\omega)$ part of α-$Al_2O_3$ vs. incident photon energy [31].

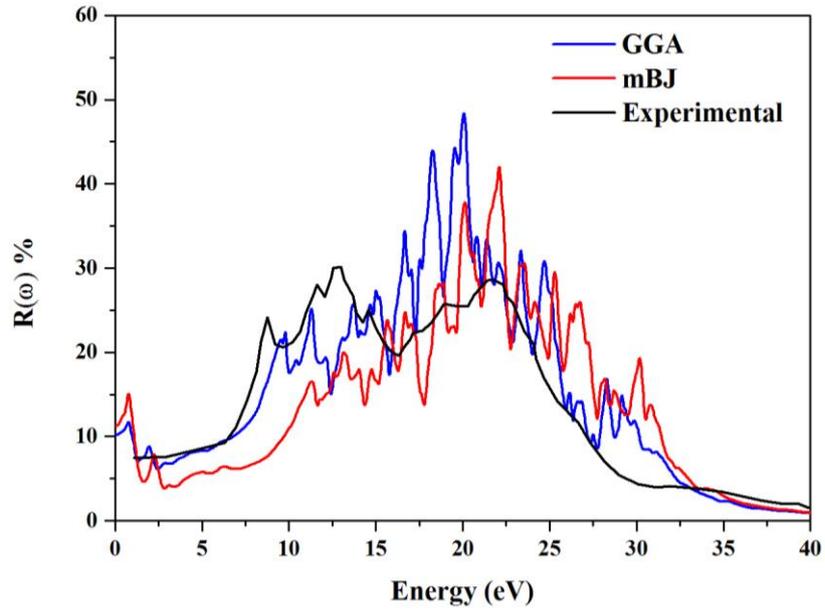

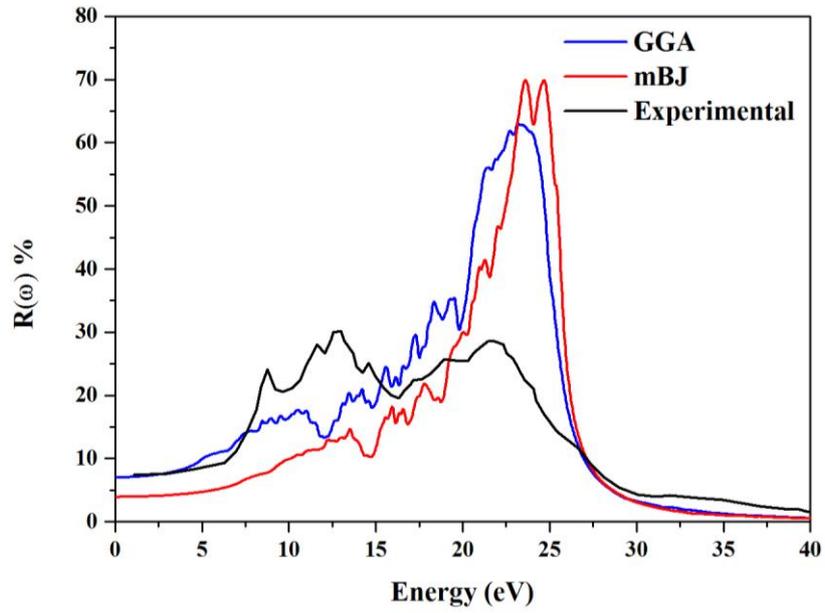

**Figure 5:** Reflectivity $R(\omega)$ of (a) γ-$Al_2O_3$ and (b) θ-$Al_2O_3$ as determined from GGA and mBJ potential. Solid black line represents the experimental data for α-$Al_2O_3$ [32].

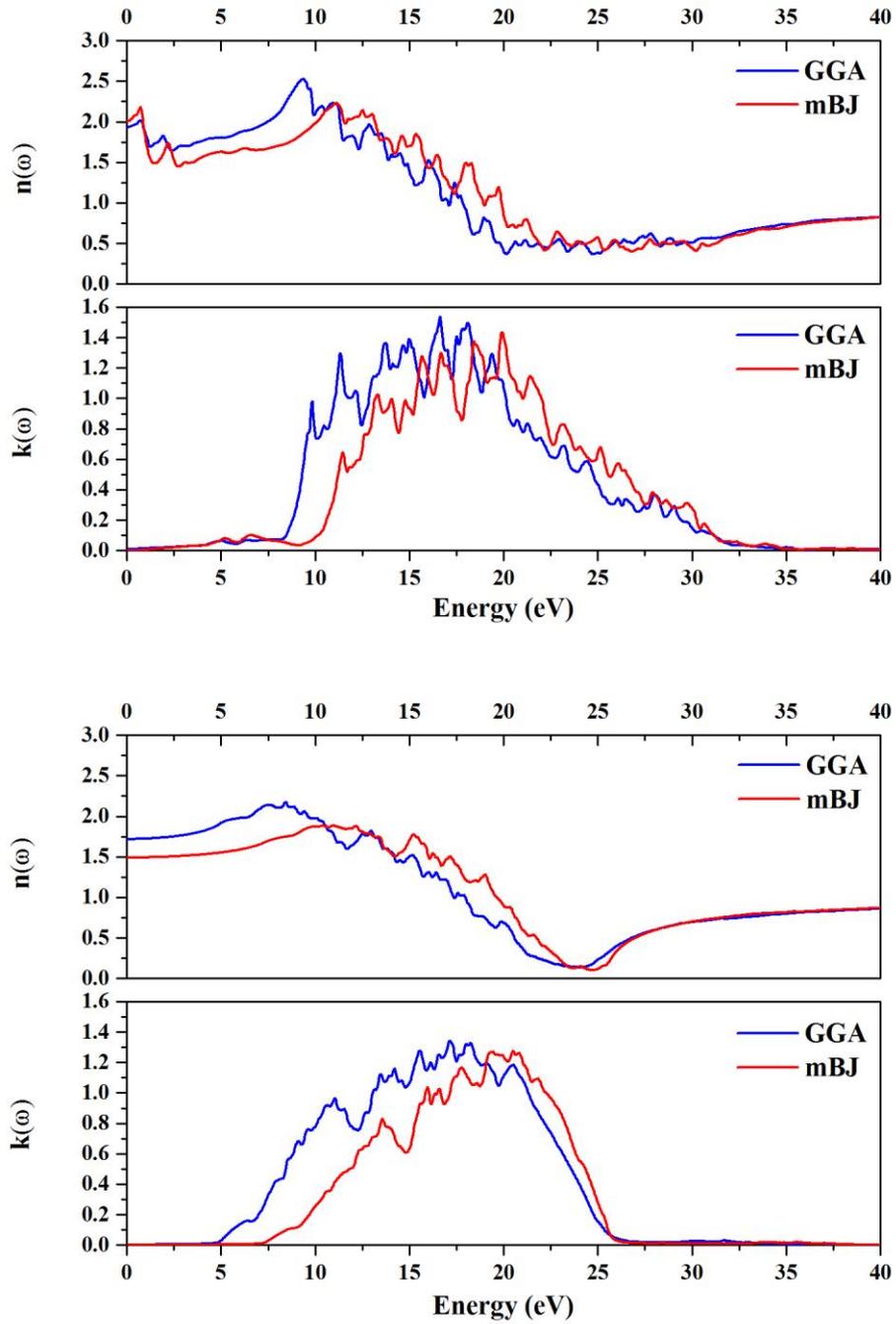

**Figure 6:** Refractive index $n(\omega)$ and extinction coefficient $k(\omega)$ for (a) γ-$Al_2O_3$ and (b) θ-$Al_2O_3$ determined from GGA and mBJ potential.

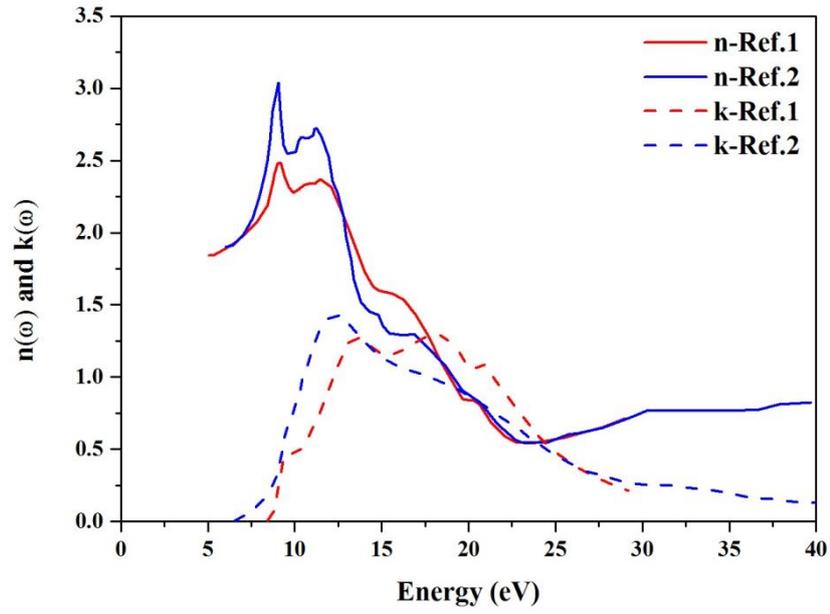

**Figure 7:** Experimental results for refractive index $n(\omega)$ (solid line) and extinction coefficient $k(\omega)$ (dashed line) of single crystal α-$Al_2O_3$, which have reported previously by Ref.1 [31] and Ref.2 [32].

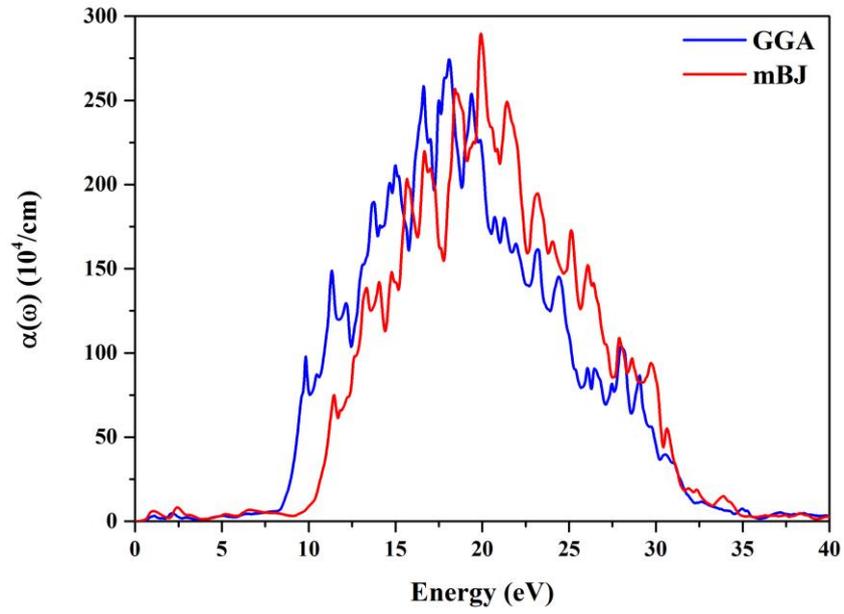

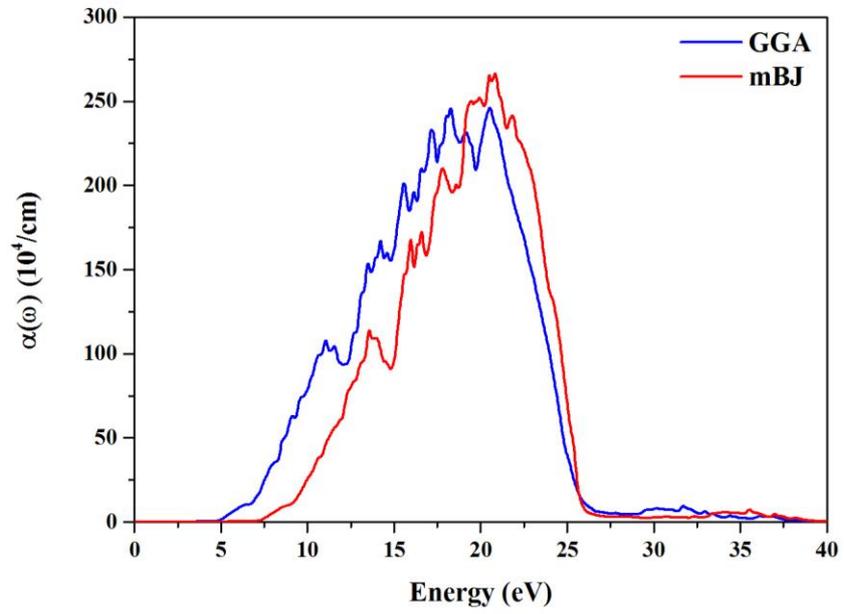

**Figure 8:** Absorption coefficients $\alpha(\omega)$ for (a) γ-$Al_2O_3$ and (b) θ-$Al_2O_3$ determined from GGA and mBJ potential.